\documentclass{acm_proc_article-sp}
\usepackage{url}
\usepackage{times}
\usepackage{graphics}
\usepackage[usenames,dvipsnames]{color}
\usepackage{listings}
\lstdefinestyle{lorenz}{%
frame=tb,
numberstyle=\tiny,
stepnumber=100,
numbersep=5pt,
fontadjust,
basicstyle=\ttfamily\small,
columns=fullflexible,
texcl,                          
tabsize=2,
showspaces=false,
showstringspaces=false,
showtabs=false,
morestring=[b]",
}
\lstdefinelanguage{aspectj}{%
keywords={thisJoinPoint,thisJoinPointStaticPart,returning,aspect,void,after,class,expected,public,protected,private,return,this,target,extends,import,around,abstract,pointcut,before,interface,declare,parents,implements,call,args,while,if,else,throws,throw,new,warning,error,perthis,pertarget,percflow,package,within,issingleton,percflowbelow,cflow,cflowbelow,thisEnclosingJoinPointStaticPart,thisEnclosingJoinPoint,get,set,for,static,execution,dominates,proceed,privileged,aspectOf},
style=lorenz,
}
\lstdefinelanguage{contractaj}{%
keywords={thisJoinPoint,thisJoinPointStaticPart,returning,aspect,void,after,class,expected,public,protected,private,return,this,target,extends,import,around,abstract,pointcut,before,interface,declare,parents,implements,call,args,while,if,else,throws,throw,new,warning,error,perthis,pertarget,percflow,package,within,issingleton,percflowbelow,cflow,cflowbelow,thisEnclosingJoinPointStaticPart,thisEnclosingJoinPoint,get,set,for,static,execution,dominates,proceed,privileged,aspectOf,@pre,@post,observer,extension,obedient},
style=lorenz,
}

\newcommand{\ajc}{AspectJ}
\newcommand{\cspace}{\text{\hspace{0.2cm}}}
\newcommand{\ncline}{\begin{array}{ll}\cspace}
  \newcommand{\ecline}{\end{array}}
  \newcommand{\conaj}{\textsc{Cona}}

  \newcommand{\PARAGRAPH}[1]{\textbf{#1.}}

  \newcommand{\beforePre}{\alpha^{\mathrm{bef}}_{\mathrm{pre}}}
  \newcommand{\beforePost}{\alpha^{\mathrm{bef}}_{\mathrm{post}}}
  \newcommand{\afterPre}{\alpha^{\mathrm{after}}_{\mathrm{pre}}}
  \newcommand{\afterPost}{\alpha^{\mathrm{after}}_{\mathrm{post}}}

\newcommand{\Ref}[2]{#1~\ref{#1:#2}}

\newcommand{\Sref}[1]{\Ref{Section}{#1}}

\author{{David~H. Lorenz} \quad {Therapon Skotiniotis}
     \\{Technical Report NU-CCIS-04-14}
     \\{College of Computer \& Information Science}
     \\{Northeastern University}
     \\{Boston, Massachusetts 02115 USA}\\
	\texttt{\{ lorenz, skotthe \}@ccs.neu.edu}
}
\title{Extending Design by Contract for Aspect-Oriented Programming}
\begin{document}
\maketitle
\pagenumbering{arabic} 

\begin{abstract}
Design by Contract (DbC) and runtime enforcement of
program assertions enables the construction of more robust
software. It also enables the assignment of blame in error
reporting. Unfortunately, there is no support for runtime
contract enforcement and blame assignment for Aspect-Oriented
Programming~(AOP).  Extending DbC to also cover aspects brings
forward a plethora of issues related to the correct order of
assertion validation.  We show that there is no generally correct
execution sequence of object assertions and aspect assertions.
A further classification of aspects as \emph{agnostic},
\emph{obedient}, or \emph{rebellious} defines the order of
assertion validation that needs to be followed.  We describe the
application of this classification in a prototyped DbC tool for
AOP named {\conaj}, where aspects are used for implementing contracts,
and contracts are used for enforcing assertions on aspects.
\end{abstract}

\section{Introduction}\label{Section:intro}

\emph{Design by Contract} (DbC)~\cite{Meyer:1992:DBC} is a
methodology for software construction that is based on runtime
enforcement of assertions.  Several object-oriented programming
(OOP) languages follow the Eiffel~\cite{Meyer:1992:Book}
example in providing support for DbC (including, e.g.,
Blue~\cite{blue} and Sather~\cite{Satherl}).  Unfortunately,
no aspect-oriented programming (AOP) language offers support
for DbC.  This paper extends DbC for controlling also the
interactions between advice and methods~\cite{Rinard:2004:CSA},
a need that is evident in any non-trivial AOP application
development~\cite{Kersten:1999:ACS}.

While runtime contract enforcement and blame assignment for
objects is well understood, it is unclear how DbC extends to
aspects.  In DbC for OOP, assertions are enforced during method
invocation; a failure clearly implicates one of two distinct
objects, the caller or the callee.  In DbC for AOP, there are
two kinds of entities (objects and aspects), two different
kinds of assertions (assertions for objects and assertions
for aspects), and no implicit caller (for an aspect's advice).

DbC for OOP also validates logical implications
between supertype assertions and subtypes assertions on
methods~\cite{Findler:2001:CSO}: an overriding method must be a
behavioral substitute~\cite{Liskov:1994:BNS} for its overridden
counterpart.  In comparison, DbC for AOP must validate that
the method with an advice is a behavioral substitute for its
advice-less counterpart.

These differences brings about several issues:
\begin{itemize}
	\item[(\emph{i})]~At what point during the
execution of the program should each kind of assertion be
checked?
	\item[(\emph{ii})]~Should there be a connection between
assertions on methods and assertions on advice and how should
that be enforced at runtime?
	\item[(\emph{iii})]~How is blame
assignment affected?
\end{itemize}

In this paper we extend the classic DbC
runtime contract enforcement mechanism to cover
AspectJ's~\cite{Lopes:1998:RDA,Kiczales:2001:OAJ,colyer:aosdbook05}
advice definitions.  We concentrate on the impact of the
relative interleaving order of object contract checking and
aspect contract checking.

In \Sref{motivation}, we show that there is no generally
applicable correct order.  In \Sref{classification}, we
develop a classification of aspects according to the way
they influence object contracts.  We classify each aspect as
either \emph{agnostic}, \emph{obedient}, or \emph{rebellious},
and show that the membership of an aspect in one of the
defined categories implies a particular order.  Based on this
classification, the execution order of method invocation (and
its advice and assertions) changes, in order to properly assign
blame for any contract violations that may occur.

Enforcing contracts via aspects is also an application
area of aspects that serves as an illustration for
the need to differentiate between agnostic, obedient,
and rebellious aspects.  In \Sref{casestudy}, we present
{\conaj}~\cite{Lorenz:2003:CAA,Skotiniotis:2004:FAB,Skotiniotis:2004:GCA,ref:cona},
a tool for the provision and enforcement of DbC in both OOP
and AOP.

\section{Motivating Example}\label{Section:motivation}

\begin{figure*}[tb]
	\centerline{%
    \input{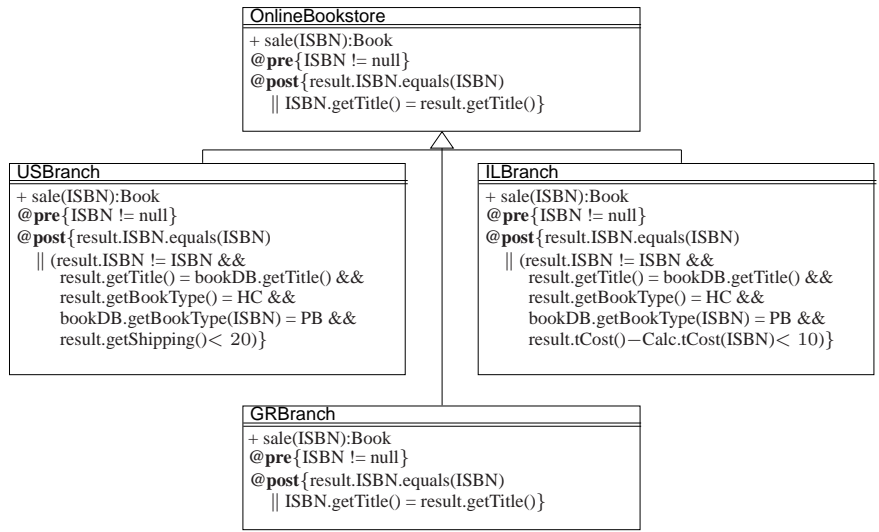}
	}
    \caption{The online bookstore has offices in USA, Greece, and Israel. Each 
    site has different taxes (Sales, VAT) and also different shipping agencies.
    \textsf{sale} is overwritten, however, the contracts associated with each 
    method implementation make the three subclasses proper behavioral subtypes of
    \textsf{OnlineBookstore}.}
    \label{example:UML}
  \end{figure*}

Consider a software system for an online bookstore
(Figure~\ref{example:UML}) with offices in the USA, Greece,
and Israel.  A book sale transaction requires a non-empty
ISBN number (pre-condition of \textsf{OnlineBookstore.sale}).
The sale completes by providing a book that either matches the
requested ISBN number or has the same title (post-condition
of \textsf{OnlineBookstore.sale}) but possibly a different
ISBN. The post-condition in \textsf{OnlineBookstore.sale}
permits to substitute the requested book with a different
edition of the book\footnote{We assume that two books with
the same title are either the same or different editions
of the same book. That is, there do not exist two books of
different contents with the same title in the bookstore.}
(e.g., a paperback (PB) version instead of a hardcover (HC)
version, or vice versa).

Concrete subclasses of \textsf{OnlineBookstore} specialize
\textsf{sale} to reflect the policy in effect in each of the three
different countries.  Specifically,
	\begin{itemize}
		\item In the USA, an order for a paperback version
			of a book that is not available in the bookstore may
			be fulfilled with a hardcover version of the book as
			long as the shipping costs for the hardcover version
			does not exceed the amount of 20 Dollars.

		\item In Greece, an order for a book that is not
			in-stock but another version (paperback or hardcover)
			is available is fulfilled by providing that version
			instead.

		\item In Israel, if the requested book is a paperback
			but the bookstore has only the hardcover version, then
			the hardcover is provided as long as the difference
			in the total cost (book price \textit{plus} shipping)
			is less than 10 Shekels.

	\end{itemize}

The post-condition on the specialized implementations of
\textsf{sale} capture the corresponding country's policy,
which needs to be checked at runtime.

In terms of contracts (transactions) the role of provider
(server) is played by the online bookstore software. The
consumer (client) role is played by the customer that uses
the online bookstore software.

\subsection{Contracts in the Presence of Aspects}
\label{section:contractsalongaspects}

Interesting issues arise in the situations where user aspects
are present in the system where contracts (regardless of
how they are being implemented) are used.  Since an aspect
may observe or alter information before, after, or around a
method's call/execution, a user aspect's advice might:
\begin{itemize}
  \item 
    Break a method's pre-condition even when the client calls the method
    correctly.
  \item 
    Break a method's post-condition even when the method's pre- and
    post-condition where fulfilled by the method's implementation.
  \item 
    Correct a call to a method $m$ that originally violated $m$'s pre-condition.
  \item 
    Correct a previously erroneous implementation that did not fulfill its
    post-condition.
  \item
    Add extra behavior to a method's implementation without altering
    the set of states accepted by the pre- and post-condition
    assertions, i.e., provide a different mapping for the same input and
    output value sets of the method.
  \item 
    Add extra behavior by extending the method's specification (pre- and/or post-conditions).
  \item 
    Monitor a method's execution by collecting information or checking certain system
    properties without affecting the behavior of the method.
\end{itemize}

The order of execution amongst aspect advice, pre- and
post-condition validation, and method execution determines
which one of the above situations occur.  For example,
behavioral extension via aspect advice should only be allowed
when aspect advice is executed before a method's pre-condition
or post-condition validation.

Consider (the addition of an aspect that will implement)
an increase of 10\% in shipping cost on hardcover books (Listing~\ref{listing:breakContract}).
\begin{figure}[tb]
\lstinputlisting%
[language=contractaj,
caption={ShippingCost aspect adding the 10\% extra shipping cost on hardcover editions.},
numberstyle=\tiny,
numbers=left,
stepnumber=1,
breaklines=true,
label=listing:breakContract,
basicstyle=\footnotesize ]
{Code/ShippingCostNoDBC.aj}
\end{figure}
\textsf{ShippingCost} is added to the system and attached to \textsf{sale}
method calls with an after advice holding the relevant code for the
enforcement of the extra cost.  Focusing on \textsf{sale} method calls,
it is not obvious what execution sequence amongst pre- and post-condition
validation, aspect advice and method execution should be followed.
Three possibilities (Table~\ref{table:executions}) are possible and
we will examine each one in turn within the context of the on line
bookstore example.

Assuming an execution policy where user aspects are executed
before pre-conditions and after post-condition are enforced
(execution order A in Table~\ref{table:executions}), clients
to both the \textsf{USBranch} and \textsf{ILBranch} may be
charged more than what was agreed.  Although the client has
maintained the originally agreed upon contract, the final
outcome breaks the clients expectations blaming the provider.

The shipping cost increase introduced via the aspect can result
to shipments where the total shipping cost will be more than
\$20 and a total cost that is more than 10 Shekels. From the
clients point of view, it is clearly an error of the online
bookstore since the agreed policy for replacing a paperback
edition was not followed.\footnote{In fact an OOP runtime
contract enforcement mechanism will blame the corresponding
bookstore branch since it does not take into account aspects
and their specification.} At the same time the corresponding
branches have assurances (\textsf{sale}'s post-condition)
that the replacement policy was honored.  DbC mechanisms
fail to correctly assign blame in the presence of aspects,
misguiding developers and increasing the time spend on bug
detection and correction.  It is clear that the party that is
actually at fault here is the aspect. One can conclude that
the aspect is at fault only after observing all three entities,
their execution sequence and their interactions.

However, an aspect can also be used to bring about the
exact opposite effect. An aspect can intervene after the
return of a method that originally returned a faulty result
which violated the post-condition and modify the result in
such a way so that now the post-condition is not violated.
The execution interleaving would have to allow such aspects to
intervene before the runtime check for a methods post-condition
(execution order B in Table~\ref{table:executions}). These
situations are examples of extensions/fixes through aspects.

In the case of the online bookstore, suppose that a 20\%
markdown is in effect for all hardcover editions. Consider
a customer in Israel ordering a book using the ISBN of the
paperback edition. Suppose the paperback edition costs 80
Shekels and the hardcover edition costs 100 Shekels. The
bookstore does not have any copies of the paperback edition
and provides the hardcover edition of the book instead. Before
applying the 20\% markdown on the hardcover edition, the
post-condition of the \textsf{ILBranch} disallows the switch
from paperback to hardcover since the difference in cost is more
than 10 Shekels.  However if the aspect is allowed to intervene
before the post-condition check is enforced then the difference
in total cost is 0 Shekels and the sale can go through.

One may resort to a conservative approach in which a
method's pre-condition is checked twice: once before
the advice and once before the method (execution order C
in Table~\ref{table:executions}).  However, that would
restrict the applicable aspects, thereby compromising
obliviousness~\cite{filman:aosdbook05}.

  \begin{table*}[tb]
    \centering{
    \begin{tabular}{|l|clclclclclclc|}
      \hline
      \emph{Policy} & \multicolumn{13}{c|}{ \emph{Execution Interleaving}} \\ \hline \hline
      \textit{Exec Order A} & $\alpha$ &$\leadsto$&$m_{\mathrm{pre}}$&$\leadsto$&$m$&$\leadsto$&$m_{\mathrm{post}}$&$\leadsto$&$\alpha$&&&&\\ \hline
      \textit{Exec Order B} & $m_{\mathrm{pre}}$&$\leadsto$&$\alpha$&$\leadsto$&$m$&$\leadsto$&$\alpha$&$\leadsto$&$m_{\mathrm{post}}$&&&&\\ \hline
      \textit{Exec Order C}& $m_{\mathrm{pre}}$&$\leadsto$&$\alpha$&$\leadsto$&$m_{\mathrm{pre}}$&$\leadsto$&$m$&$\leadsto$&$m_{\mathrm{post}}$&$\leadsto$&$\alpha$&$\leadsto$&$m_{\mathrm{post}}$ \\ \hline 
    \end{tabular}
    \caption{Three alternative policies of interleaving execution (form left to right) of contracts checking pre- and post-conditions ($m_{\mathrm{pre}}$ and $m_{\mathrm{post}}$ respectively) with advice from an aspect ($\alpha$).}\label{table:executions}
    }
  \end{table*}

\section{A Classification of Aspects for DbC}\label{Section:classification}

Aspects can be used:
\begin{itemize}
	\item[(\textit{i})]
to enforce properties without altering the behavior
of the underlying system (e.g., logging the online
bookstore system, checking the online bookstore coding
style~\cite{Lieberherr:2003:CSE}, implementing the
DbC assertions found in Figure~\ref{example:UML}
[cf.~\Sref{casestudy}]. 
	\item[(\textit{ii})] to allow for
extensions to the behavior of the underlying system,
(e.g., allowing for extra charges/discounts on the online
bookstore).
\end{itemize}
The two different uses of aspects
in the presence of contracts imposes certain restrictions on
their execution order. Furthermore, it may result in erroneous
blame assignments complicating error detection and resolution.

The choice of which interleaving execution order
(Table~\ref{table:executions}) to enforce also depends on the
mechanism of assigning blame in cases of violation.  Standard
DbC mechanisms are ignorant of aspects.  Blame assignment must
therefore be extended to deal with aspects as entities that
can be assigned blame. Also, what is being violated has to be
redefined in order to take into account the intention of code
found in aspect definitions.

The ability to define an aspect's intentions through a clear
declarative specification as well as the runtime validation
of these intentions is crucial in error detection, error
resolution and reasoning about aspect-oriented programs.
We identify three intentional categories of aspects: namely,
\emph{agnostic}, \emph{obedient}, and \emph{rebellious}.

\subsection{Agnostic Aspects}

\emph{Agnostic} aspects are aspects that do not affect
a method's assertions in any way.  Agnostic aspects
are ``sandwiched'' with the original method's pre- and
post-conditions.  From the perspective of the callers of
the method, the method with the advice (as on body) has the
same assertions as the original method body.  In addition,
assertions do not change from the method's perspective either.

\subsubsection{Execution Order}

The language imposes the following execution sequence in the
case of agnostic aspects (execution order C):
$$
\begin{array}{l}
m_{\mathrm{pre}}\leadsto  \beforePre \leadsto  \alpha \leadsto\beforePost\leadsto m_{\mathrm{pre}}\leadsto m\leadsto m_{\mathrm{post}} \leadsto\\
\leadsto\afterPre\leadsto \alpha\leadsto  \afterPost\leadsto m_{\mathrm{post}}
\end{array}
$$
Where 
\begin{itemize}
  \item 
    $\beforePre$ denotes to the pre-condition of the aspect's ($\alpha$) before advice
  \item 
    $\beforePost$ denotes to the post-condition of the aspect's ($\alpha$) before advice
  \item 
    $m_{\mathrm{pre}}$ denotes to the method's ($m$) pre-condition 
  \item 
    $m_{\mathrm{post}}$ denotes to the method's ($m$) post-condition 
  \item 
    $\afterPre$ denotes to the pre-condition of the aspect's ($\alpha$) after advice
  \item 
    $\afterPost$ denotes to the post-condition of the aspect's ($\alpha$) after advice
\end{itemize}
This execution sequence makes sure
that the original method's pre- and post-condition is validated both
before advice execution and after advice execution. 

\subsubsection{Blame Assignment}

Runtime assertion validation and blame assignment for agnostic aspects is
described in Table~\ref{table:observer-validation}. On top of checking for
each individual assertion at runtime extra implications between assertions
are validated to guarantee proper execution flow between aspect advice
and method implementation. The method's pre-condition has to imply the
before advice pre-condition making sure that the aspect developer took
into account the valid start states of the method. Furthermore, before
advice post-condition has to imply the methods pre-condition. Since
no alteration of behavior is allowed in agnostic aspects execution of
the method's implementation must start in a valid state satisfying the
methods pre-condition.

\begin{table}[b!]
  \centerline{
  \begin{tabular}{|c|c|}
    \hline
    Assertion Validation & Blame Assignment \\
    \hline
    \hline
    $m_{\mathrm{pre}}$ & Caller\\
    \hline
    $m_{\mathrm{pre}} \rightarrow \beforePre$ & Aspect \\
    \hline
    $\beforePost$ & Advice \\
    \hline
    $\beforePost \rightarrow m_{\mathrm{pre}}$ & Aspect \\
    \hline
    $m_{\mathrm{pre}}$ & Advice \\
    \hline
    $m_{\mathrm{post}}$ & Method \\
    \hline
    $m_{\mathrm{post}} \rightarrow \afterPre$ & Advice \\
    \hline
    $\afterPre$ & Advice \\
    \hline
    $\afterPost$ & Advice\\
    \hline
    $\afterPost \rightarrow m_{\mathrm{post}}$ & Aspect\\
    \hline
    $m_{\mathrm{post}}$ & Aspect \\
    \hline
  \end{tabular}
  }
  \caption{Implications for agnostic aspect assertions and blame assignment. Order of execution goes from top (first) to
  bottom (last).}
  \label{table:observer-validation}
\end{table}

Similarly, the method's post-condition must imply the after advice
pre-condition. This is again the responsibility of the aspect developer
who is required to begin any agnostic aspect after advice from a valid
state according to the method's post-condition. Finally, after advice
post-condition must directly imply the method's post-condition. Upon
completion of an agnostic aspect's after advice there should be no change in
the method's behavior and specification. Furthermore, the return values
of the after advice are the return values of the method call as a whole
and thus the method's post-condition must also hold.

\subsection{Obedient Aspects}

Another form of extension to the base system is one where
the set of input and output states remains the same but the
mapping of input to output values changes. In these situations
the aspect with such an intended extension is categorized as
an obedient aspect.

\emph{Obedient} aspects are aspects used to provide extra
behavior without changing the method's pre- and post-
conditions, i.e., obedient aspects just provide a different
mapping from the same input to the same output value sets of
the method.  From the perspective of a caller of the method,
the method with the advice has the same assertions as the
original method had.

\subsubsection{Execution Order}

Declaring an obedient aspect
imposes the following execution sequence (execution order~B):
$$ 
 m_{\mathrm{pre}} \leadsto \beforePre \leadsto \alpha \leadsto \beforePost \leadsto  m \leadsto 
 \afterPre \leadsto \alpha \leadsto \afterPost \leadsto m_{\mathrm{post}}
$$
Adding an obedient aspect does not alter the pre- and post-conditions of
the method as they are known to the rest of the system.
The first assertion validation is still the original method's
pre-condition and the last assertion validation is the original method's
post-condition.

\subsubsection{Blame Assignment}

Once the method's pre-condition has been successful validated this should
immediately imply the pre-condition for the aspect's before advice
is also true. The aspect developer knows the method's pre-condition
and has defined the aspect to be an obedient aspect, it is therefore the
aspect developer's responsibility to accept all legal starting states
of the method as legal starting states of the aspect's before
advice. Failing to do so will signal an error blaming the aspect developer
for attempting to use an obedient aspect without taking into account the
original method's pre-condition (Table~\ref{table:friend-validation}).

\begin{table}[b!]
  \centerline{
  \begin{tabular}{|c|c|}
    \hline
    Assertion Validation & Blame Assignment \\
    \hline
    \hline
    $m_{\mathrm{pre}}$ & Caller\\
    \hline
    $m_{\mathrm{pre}} \rightarrow \beforePre$ & Aspect \\
    \hline
    $\beforePost$ & Advice \\
    \hline
    $m_{\mathrm{post}} \rightarrow \afterPre$ & Aspect \\
    \hline
    $\afterPre$ & Advice \\
    \hline
    $\afterPost$ & Advice\\
    \hline
    $\afterPost \rightarrow m_{\mathrm{post}}$ & Aspect\\
    \hline
  \end{tabular}
  }
  \caption{Assertions validated for obedient aspects (execution flows from
  top to bottom) along with blame assignments.}
  \label{table:friend-validation}
\end{table}

The aspect's before advice post-condition is then checked.
Failing the before advice post-condition blames the advice code
for not fulfilling the expected assertion upon its termination.
The next assertion to be checked is an implication relation
between the original method's post-condition and the after
advice pre-condition. Blame is assigned to the aspect developer
in the case of failure, for declaring an obedient aspect
and not taking into account the method's post-condition as a
valid start state for the aspect's after advice.  The after
advice pre-condition is then checked, blaming the composition
rules (pointcuts) in the case of an error. The after advice
post-condition validation follows which blames the advice code
in case of an error. Finally an implication between the after
advice post-condition and the original method's post-condition
is validated. This check makes sure that the state in which
the after advice terminates (and thus the whole call to the
method) does so in valid state according to the method's
original post-condition.

The difference between agnostic and obedient aspects is
subtle. In both cases, if the method pre-condition is valid,
we can invoke the method with its advice, and the result
satisfies the post-condition.  However, in the case of the
obedient aspect it is possible that a before aspect will
invalidate the precondition, the method is executed on a state
that does not satisfy the precondition, and an after advice
fixes the postcondition (if necessary).  With agnostic aspects,
the method code executes only in the context it was meant for.

\subsection{Rebellious Aspects}

\emph{Rebellious} aspects are aspects used only to provide
behavioral extensions to existing methods.  Rebellious aspects
change the behavior of existing methods.  After the aspect is
applied to a method, from the perspective of a caller of the
method, the method with the advice has different assertions
than the original method had.  However, assertions do not
change from the method's perspective.

As the category name implies, these are aspects that are determined to
alter the behavior of a method to the extend where the existing pre-
and post-conditions of the method are affected. Nonetheless, rebellious
aspects can alter
a method's pre- and post-conditions in a controlled manner:
\begin{itemize}
    \item
      For all valid start states of the method's pre-condition, the new
      pre-condition (aspect's before pre-condition) has to also be valid.
    \item 
      For all valid states according to the new post-condition (aspect's
      after post-condition) the original method's post-condition has to
      also be valid.
\end{itemize}
The above implications between the extend pre-condition (and original
method pre-condition) and extended post-condition (and the original method's
post-condition) ensure proper behavioral subtype between the original
method and the extension to the method.

\subsubsection{Execution Order}

A rebellious aspect is guaranteed an execution sequence imposed
by the aspect that allows its advice to execute even before
the method's pre-condition. This enables the redefinition
of the method's pre-condition.  In order to allow for the
methods behavioral extension the execution order enforced by
the language is (from left to right):
$$ 
 \beforePre \leadsto  \alpha \leadsto \beforePost  \leadsto m_{\mathrm{pre}} \leadsto  m \leadsto   m_{\mathrm{post}}  \leadsto
 \afterPre \leadsto  \alpha \leadsto \afterPost
$$
In this way an extended version of the method can be provided to client
programs. This kind of extension, however, raises an issue with clients
as to which of the methods pre-conditions are clients supposed to follow?

The answer is either one. A rebellious aspect provides a
behavioral method extension without breaking the existing
method's contracts, i.e., allows for a broader set of valid
start states and a narrower set of valid end states. This is
reflected in the way implication between assertions are being
validated, after checking for the validity of each assertion,
extra implications between pre- and post-conditions are
further validated.

\subsubsection{Blame Assignment}

Blame assignment is described in Table~\ref{table:extension-validation}.
The extra implication $m_{\mathrm{pre}} \rightarrow \beforePre$ makes
sure that the extension made through the aspect definition maintains
the previously valid start states for the method. In case that it does
not then the blame lies with the aspect developer for providing an
extension that breaks existing clients of the method. The implication
$\beforePost \rightarrow m_{\mathrm{pre}}$ verifies that after the
termination of before advice and before the execution flows to the
method's implementation, the program state satisfies the method's
pre-condition. If before the advice post-condition implies the method's
pre-condition then the method is not being wrongly used by the aspect. The
implication $m_{\mathrm{post}} \rightarrow \afterPre$ verifies that upon
completion of the method body execution flows into the aspect's after
advice in a correct start state for the after advice.  It is therefore
the responsibility of the aspect developer to make sure that all valid
states reached upon return of the method are also valid start states
for the aspect's after advice. Finally, the implication $\afterPost
\rightarrow m_{\mathrm{post}}$ verifies that all the valid states at
after advice termination are also valid according to the original method's
post-condition. This last check verifies that the set of valid end states
is the same as, or a subset of, the original method's set of end states.

\begin{table}[h!]
  \centerline{
  \begin{tabular}{|c|c|}
    \hline
    Assertion Validation & Blame Assignment \\
    \hline
    \hline
    $m_{\mathrm{pre}} \rightarrow \beforePre$ & Aspect \\
    \hline
    $\beforePre$ & Caller \\
    \hline
    $\beforePost$ & Advice \\
    \hline
    $ \beforePost \rightarrow m_{\mathrm{pre}}$ & Aspect \\
    \hline
    $m_{\mathrm{pre}}$ & Advice \\
    \hline
    $m_{\mathrm{post}}$ & Method \\
    \hline
    $m_{\mathrm{post}} \rightarrow \afterPre$ & Aspect \\
    \hline
    $\afterPre$ & Advice \\
    \hline
    $\afterPost$ & Advice\\
    \hline
    $\afterPost  \rightarrow m_{\mathrm{post}}$ & Aspect\\
    \hline
  \end{tabular}
  }
  \caption{Implications for rebellious aspect assertions and blame assignment. Order of execution goes from top (first) to
  bottom (last).}
  \label{table:extension-validation}
\end{table}

\subsection{Application and Obliviousness}

A DbC for AOP system is able to provide for the runtime
validation of assertions on both aspects and objects allowing
for the detection of errors due to erroneous aspect advice
implementations, erroneous aspect composition and aspect
category violations allowing for faster error detection and
resolution.

The categorization into \emph{agnostic}, \emph{obedient},
and \emph{rebellious} covers all behaviors that aspects can
introduce to a system.	Along with the incorporation of pre-
and post-conditions for advice, DbC for AOP can resolve the
execution order question by, e.g., introducing three new
keywords to the AspectJ language that programmers can use
to declare the categorizations of aspect implementations.
DbC support for AspectJ would then use this classification
to enforce the appropriate execution sequence according to
each category.	AspectJ could also extend DbC to aspects
by allowing pre- and post-condition validation for advice
\textit{and} their implication on pre- and post-conditions on
their attached methods.  Such an implementation of DbC for AOP
does not impose any restrictions on the base code, allowing
the same level of obliviousness~\cite{filman:aosdbook05}
as in the current implementations of AspectJ.

\section{A Case Study in Enforcing Contracts using Aspects with {\large\textsc{CONA}}}\label{Section:casestudy}

So far we tacitly ignored the question of how the
enforcement of assertions on methods (and on advice)
is implemented.  Enforcing runtime contract checking in
a program is a classical cross-cutting concern by its
very nature of monitoring calls made between software
entities~\cite{con-aspect-patent,Constantinides:2002:RAC}.
At the heart of any runtime contract checking tool lies the
ability to observe and intervene between calls made from one
software unit to another. The kinds of checks that need to be
carried out between calls differ depending on the paradigm
(e.g., functional~\cite{Findler:2002:CHO} vs. OO etc.).
The ability of AOP technologies to non-invasively extend
a system, along with the ability to encapsulate and compose
cross-cutting concerns, make AOP an ideal implementation
tool for DbC.

\textsc{Cona}~\cite{Skotiniotis:2004:Cona} is a DbC tool for both
OOP and AOP.  It extends Java's and AspectJ's syntax with
contracts and enforces their runtime validation.  {\conaj} works
by generating \ajc\ aspects to enforce the runtime contract
checking.  {\conaj} generates aspect definitions only for the
validation of contracts on Java classes; it generates class
definitions for the validation of contracts on AspectJ aspects.

In a {\conaj} application there
are always two ``kinds'' of aspects
\begin{enumerate}
  \item 
    \textit{User-Aspects} these are aspect definitions provided by developers that affect the underlying 
    Java program.
  \item 
    \textit{Contract-Aspects} these are generated by {\conaj} and are responsible for the enforcement of contracts for objects (DbC for OOP).
\end{enumerate}
and two ``kinds'' of objects
\begin{enumerate}
  \item 
    \textit{User-Objects} these are all the instances of user defined Java classes.
  \item 
    \textit{Contract-Objects} these are all the instances of the {\conaj} generated Java classes used for the runtime validation of
    contracts defined inside user aspects.
\end{enumerate}

Contracts on objects are implemented by contract aspects;
contracts on aspects are implemented by contract objects.
The desired DbC functionality is guaranteed provided that the
generated contract aspects are the only aspects in the system.
However, in the presence of user aspects, and in particular, the
interplay between a contract aspect and user aspects, enforcing
assertions correctly is non-trivial.  {\conaj} is therefore
an interesting testbed for exposing the issues of execution
order and a natural client for our aspect classification.

In the remaining part of this section we further explain
{\conaj} and the language extensions to Java and AspectJ as
well as \conaj's mechanisms for contract enforcement. First
an overview of DbC for Java is presented followed by a more
detailed analysis of how {\conaj} extends DbC for OOP to achieve
DbC for AOP.

\subsection{An AOP Solution to DbC for OOP}

\begin{figure}[b!]
\lstinputlisting%
[language=aspectj,
caption={An aspect definition generated by \textsc{Cona} for enforcing%
contract obligations of \textsf{GRBranch}.}, 
frame=tb,
numbers=left,
basicstyle=\footnotesize,%
numberstyle=\tiny,
stepnumber=1,
breaklines=true,%
escapechar={^},%
label=aspectcontract]%
{Code/GRBranch_Contract.aj}%
\end{figure}

  \begin{table}[b]
    \centering{
    \begin{tabular}{|c|c|}
      \hline
      Contract Value & Blame Assignment \\
      \hline
      \hline
      $\neg m_{\mathrm{pre}}$ & Caller \\ \hline
      $\forall \tau,\tau' \ (\tau' \!\prec: \tau \wedge
      \tau\!\!::m \wedge \tau'\!\!::m) \rightarrow$ & \\$ \neg (\tau\!\!::\!\!m_{\mathrm{pre}} \rightarrow
      \tau'\!\!::\!\!m_{\mathrm{pre}})$ & $\tau'$ \\ \hline
      $\neg m_{\mathrm{post}}$ & Callee \\ \hline
      $\forall \tau,\tau'\  (\tau' \! \prec: \tau \wedge \tau\!\!::m
      \wedge \tau'\!\!::m) \rightarrow$ & \\$ \neg (\tau'\!\!::\!\!m_{\mathrm{post}} \rightarrow
      \tau\!\!::\!\!m_{\mathrm{post}})$ & $\tau'$ \\ \hline
    \end{tabular}
    }
    \caption{Blame assignment rules for OOP.  $\tau'\prec:\tau$
    defines that $\tau'$ is a {direct} subtype of
    $\tau$. $\tau\!\!::\!\!m_{\textrm{pre}}$ denotes the method $m$
    with pre-conditions $\textrm{pre}$ is defined in type $\tau$.}
    \label{table:oo-dbc-rules}
  \end{table}

  \begin{figure*}[t!]
	\centerline{%
	\input{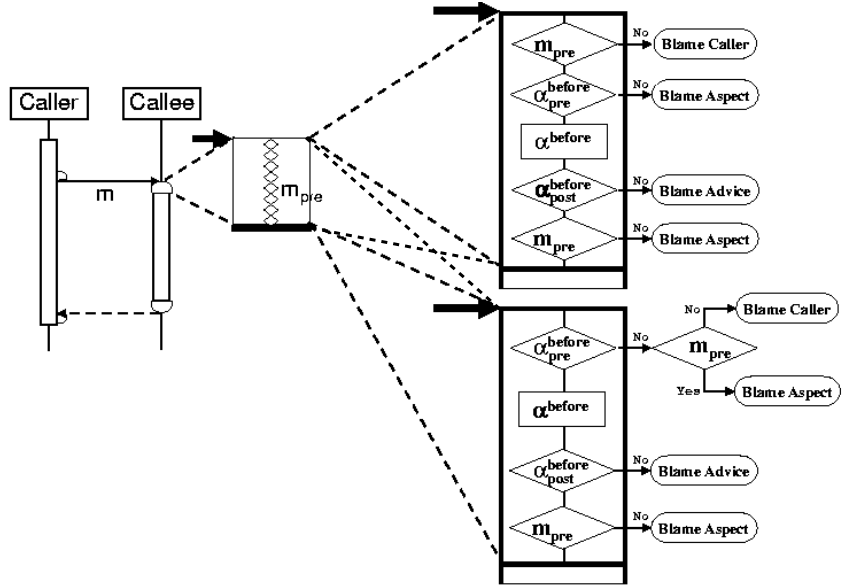}
    }
    \caption{A flowchart diagram superimposed on
a sequence diagram for a method call to $m$ showing the order of contract evaluation for obedient aspects (top) and
    rebellious aspects (bottom). $m_\mathrm{pre}$ refers to $m$'s pre-conditions assertion and $\alpha^\mathrm{before}_\mathrm{pre}$ and $\alpha^\mathrm{before}_\mathrm{post}$ to the aspect's pre- and post-condition assertions respectively.}
  \label{sequence}
  \end{figure*}

The AOP implementation for the provision of DbC for OOP maps
each Java type to a contract aspect that is responsible for
the enforcement of all contracts defined in that type. Pre-
and post-conditions and invariant expressions are generated as
aspect methods. Pointcuts are generated to capture executions
of a type's methods. Pointcuts are further used to distinguish
which type's method is called and which type's implementation
of this method is actually executed. In this way the correct
type is blamed in the case of a contract violation. Extra
methods are generated which recursively traverse contracts
of the current type's supertypes verifying the correctness of
the type hierarchy.

Listing~\ref{aspectcontract} shows an example of a contract
aspect implementing a contract in AspectJ.  The code has been
generated by \textsc{Cona}~\cite{Skotiniotis:2004:Cona} and
it illustrates that aspects implementing contracts can follow
a pattern (template): pointcut definitions capture calls
to the method (lines~\ref{pcd:begin}--\ref{pcd:end}),
before advice checks method pre-conditions
(lines~\ref{before:begin}--\ref{before:end}),
and after advice checks method's post-conditions
(lines~\ref{after:begin}--\ref{after:end}). Invariant
assertions are checked both before and after a call
to the object's public methods.  Auxiliary methods are
generated inside aspects to deal with hierarchy checking
(lines~\ref{aux:begin}--\ref{aux:end})~\cite{Findler:2001:CSO}.
The recipe for \textsc{Cona}'s aspect generation is explained
elsewhere~\cite{Skotiniotis:2004:GCA}.

Blame assignment for OOP~\cite{Findler:2001:CSO} is summarized
in Table~\ref{table:oo-dbc-rules}. The last two rows display
the type hierarchy check performed in order to verify
proper behavioral subtyping.  Due to the lack of aspectual
polymorphism~\cite{Ernst:2001:AP,Ernst:2003:APA} in AspectJ, the traversal
of contracts in {\conaj} deploys AspectJ's reflective features
in order to acquire aspect instances at runtime.

The AOP implementation of DbC for OOP works well as long as
the generated aspects are the only aspects in the system.

\subsection{An OOP Solution to DbC for AOP}

A DbC for AOP solution has to be able to enforce the execution
sequence required by each aspect category as well as manage pre-
and post-conditions found in user aspect definitions.

Figure~\ref{sequence} uses a flowchart diagram superimposed
on an interaction diagram showing when obedient (top) and a
rebellious (bottom) aspects intervene during $m$'s execution,
and what is the order of contract evaluation for these two
aspect categories.\footnote{Agnostic aspects are similar
to obedient aspects with some additional contract checks
interleaved during execution.} Aspect category definitions
denote a specific order of contract evaluation as well as
implications between assertions in contract aspects and
contract objects. All diamond shaped decisions points are
generated by \conaj. Decisions points denoted as diamonds in
Figure~\ref{sequence} and labeled with an $\alpha$ pattern are
captured as Java classes (contract objects). Diamond shaped
decision points labeled with an $m$ pattern are captured as
AspectJ aspects (contract aspects).

To support pre- and post-conditions for before and after
advice we assume the AspectJ language was extended with
the keywords \texttt{agnostic}, \texttt{obedient} (e.g.,
Listing \ref{inputsample}, line 1), and \texttt{rebellious},
and, furthermore, that advice definitions can be
annotated with pre- and post-conditions that apply to
the corresponding advice body (Listing~\ref{inputsample}
lines~\ref{predef}--~\ref{postdef}). Pre- and post-conditions
are comprised of side effect free boolean expressions.

\begin{figure}
\lstinputlisting%
[
language=contractaj,
caption=An example of using pre- and post-conditions in the ShippingCost aspect,
breaklines=true,
escapechar={^},
label=inputsample,
numbers=left,
stepnumber=1,
numberstyle=\tiny,
basicstyle=\footnotesize ]
{Code/ShippingCost.aj}
\end{figure}

The extension of {\conaj} for adding contracts to
advice takes as input aspect definitions with pre- and
post-conditions. Through a preprocessing stage, new aspects are
generated to verify contracts on object instances \emph{and}
auxiliary class definitions to handle pre- and post-conditions
on user defined advice.

First we present a sample program definition of advice
with pre- and post-conditions along with the generated
output.  Listing~\ref{inputsample} shows the obedient
\textsf{ShippingCost} aspect presented earlier for the online
bookstore example written in \conaj.
Listing~\ref{sampleoutput}
shows the same aspect as Listing~\ref{inputsample} after
being processed by \conaj.  The auxiliary class definitions
(Listings~\ref{pregeneratedclass} and~\ref{postgeneratedclass})
enforce the implications between contract objects and contract
aspects.

\begin{figure}[t!]
\lstinputlisting%
[
language=aspectj,
caption=Sample output for \texttt{ShippingCost} aspect showing the wrapping of advice to check pre- and post-conditions,
frame=tb ,
numbers=left,
breaklines=true,
numberstyle=\tiny,
basicstyle=\footnotesize,
stepnumber=1,
escapechar={^},
label=sampleoutput]
{Code/Contract_ShippingCost.aj}
\end{figure}

\begin{figure}
\lstinputlisting%
[
language=contractaj,
caption=Pre-condition wrapper for aspect after advice,
breaklines=true,
frame=tb,
numbers=left, 
basicstyle=\footnotesize,
numberstyle=\tiny,
stepnumber=1,
escapechar={^},
label=pregeneratedclass ]
{Code/Contract_ShippingCost_After_PreCond.java}
\end{figure}

\begin{figure}[t!]
\lstinputlisting%
[
language=contractaj,
caption=Post-condition wrapper for aspect after advice,
breaklines=true,
frame=tb,
numbers=left, 
basicstyle=\footnotesize,
numberstyle=\tiny,
stepnumber=1,
escapechar={^},
label=postgeneratedclass ]
{Code/Contract_ShippingCost_After_PostCond.java}
\end{figure}

Pre- and post-conditions inside advice must be side effect
free Java boolean expressions. These expressions can refer to
values that relate to the state of the aspect, and also to the
state of the receiver and the caller of method invocations.
All information concerning a join point (e.g. \texttt{target},
\texttt{args}, \texttt{source}, etc.) can be referred to from
inside an aspect's pre- and post-conditions definitions.

The programmer must specify the aspect's category and
the acceptable states in which advice may start/finish.
Failing to meet the pre-condition of a block of advice
implies that the attachment of the specific aspect to the
base program $P$ is not correct (Listing~\ref{sampleoutput}
line~\ref{failedpre}) and the aspect gets blamed.  Similarly, if
the post-condition of a piece of advice fails, this implies that
the code inside the advice did not meet up to its obligations
(Listing~\ref{sampleoutput}, line~\ref{failedpost}) and the
advice code gets the blame.

Once pre- and post-conditions have been satisfied, the way
by which pre- and post-conditions of advice interplay with
method pre- and post-conditions of methods that they advice are
checked for correctness based on the aspects defined category.

In the case of an obedient aspect (i.e., \textsf{ShippingCost}),
before the execution of the after advice block the implication
between the method's post-condition and the after advice
pre-condition (Listing~\ref{sampleoutput},
line~\ref{implication1}) is checked by passing control to an auxiliary
class (Listing~\ref{pregeneratedclass}). Failing to satisfy
this implication throws an \textsf{ObedientViolation}
(Listing~\ref{sampleoutput}, line~\ref{fviolated}) exception
pointing out the erroneous implementation of the aspect's
before advice.  The advice pre-condition and post-condition is
then checked, blaming the aspect's composition and the after
advice implementation respectively (Listing~\ref{sampleoutput},
lines~\ref{failedpre} and~\ref{failedpost})

\section{Discussion and Related Work}\label{section:relatedandfuture}

We have extended DbC for AOP to handle the interaction between
pre- and post- conditions of methods and advice.  Our goal
was primarily to demonstrate the feasibility a novel approach
for supporting DbC for AOP without compromising obliviousness.
We have also presented a prototyped tool for enforcing DbC by
generating aspects.  This work paves the road to extending DbC
for AOP but also leaves for future work a few issues that are
need to be worked out for a complete practical support of DbC
for AOP.

\PARAGRAPH{Invariants and Object State}
DbC makes the
obligation--benefit contract between software consumers
and providers explicit.  Each instance method defines
the valid states in which its execution can start
(\emph{precondition}), and the states in which it may
terminate (\emph{postcondition}). A more general assertion
(\emph{invariant}), which is maintained before as well as after
any externally observable state of an object, ensures that the
object maintains an acceptable state throughout the program's
execution.  Object and aspect invariants can be checked
by public methods as part of the pre- and post-conditions
using the same execution order we have established.  However,
to handle invariants, DbC for AOP needs not only control the
interaction between aspects and object behavior, but also
between aspects and object state.  It is also not enough to
monitor only call and execution join points.  Since advise can
change the state of an object, the interaction between advice
and objects' state need to checked to enforce invariants.
Additional checks are needed for checking the invariant of
super-classes and outer classes.  Invariants for aspects are
particularly useful as an inductive hypothesis: What ever is
assumed should be inductively provable.

\PARAGRAPH{Other Kinds of Advice} We have focused on the
interaction between methods and before and after advice. An
around advice replaces the method entirely and may or may
not contain a proceed() call in its body. Such a narrowing
or replacement interaction~\cite{Rinard:2004:CSA} should be
handled similar to how overriding methods handle a call to super.
One way of interpreting the aspect annotation for around advice
is to require that an around advice of the form $f();proceed()$
would behave like a before advice would, and an around advice
of the form $proceed();f()$ would behave like an after advice.

\PARAGRAPH{Introductions and Pointcut Descriptors}
We have concerned ourselves with the pointcut and advice mechanism
in AspectJ.  AspectJ also support a static OC mechanism (e.g.,
introductions).  This is yet another form for a possible interaction
between an aspect and a class which we do not handle.  Another subtle
issue is the interaction between aspects and pointcuts.  In our support of
DbC for AOP we have thus far assumed that pointcuts have no side effects.
This, however, is not generally true, not even for AspectJ.  One might
consider augmenting pointcut descriptors with pre- and post-conditions
of their own.

\PARAGRAPH{Aspect Composition Validation Tool}
Klaeren et al.~\cite{Klaeren:2000:ACA} present the \emph{Aspect
Composition Validation Tool} for checking pre- and post-conditions
for aspect compositions according to configurations of the system's
components.  The tool is developed using an older version of \ajc\
(0.4beta7) which is drastically different than version 1.0.6.
Further more, a set of configuration rules is added through
\ajc's introductions and composition is validated according to these
rules. Correctness is defined to be a valid aspect configuration that will
allow a receiver to perform its task as specified by the overall system
specification. Unfortunately, checking that the behavior of attached
aspects and the base system is well defined is not verified. As long
as the composition of aspects is within the set of valid compositions,
the system is correct. An aspect can therefore break a methods pre-
or post-condition as long as the configuration at hand allows it. There
is no clear distinction between a type's obligations in their system,
since the same call to an instance of the same type can behave differently
depending on its aspect configuration.

\PARAGRAPH{JML}
Clifton and Leavens~\cite{Clifton:2002:OAP} use behavioral specifications
for aspects as a means to assist in modular reasoning for aspect-oriented
programs. Aspects are categorized as either observers that do not
alter the behavioral specification of their attached methods, or as
assistants that can alter behavior. Their categorization was a result
after inspecting available AspectJ code and from discussions within the
aspect community.

Assertions on aspects as well as objects are expressed in the Java
Modeling Language (JML)~\cite{Leavens:2000:JML}. The AspectJ syntax
is extended
with three main features:
\begin{enumerate}
  \item 
    JML can be used inside aspect definitions, defining the aspect's
    behavioral specification
  \item 
    Keywords \texttt{observer} and \texttt{assistant} can be used to
    annotate aspect definitions with their expected intend in the system.
  \item 
    The statement \texttt{accept(}\textit{TypePatern}\texttt{)} has
    to be used inside modules, making explicit the modules intent to
    allow assistance from \textit{TypePatern}.
\end{enumerate}
The specification of a method along with its assistant is created as
a graph by following the possible execution paths, logically and-ing
assertions and binding parameter variables and model variables to values.

Although our goals are different than those of Clifton and Leavens the
proposals share some ideas. Our notion of aspect categorization comes from
an analysis of how aspects can affect a program execution and not through
an analysis of existing usage of aspect oriented programming.  In doing
so we aim at providing a categorization that covers any possible usage
of aspects rather than the common usage of aspects. Further more, our
proposal brings these categorizations into the programming language and
enforces an execution order disallowing aspect behavior outside the bounds
of its defined category. Enforcing an execution sequence helps in defining
the semantics of assertion validation along with blame assignment without
having to resort to long and complicated path specifications based on
control flow paths. Finally, in our proposal modules are not affected in
any way and remain oblivious to the addition of aspects. Obliviousness is
decreased in Clifton et al.~\cite{Clifton:2002:OAP} with the incorporation
of the \texttt{accept} expression in the language.

\PARAGRAPH{Pipa}
Pipa~\cite{Zhao:2003:Pipa} defines a Behavioral Interface Specification
Language (BISL) tailored for \ajc\ along the ideas of Clifton et
al.~\cite{Clifton:2002:OAP}. Pipa statements extend the Java Modeling
Language (JML) to accommodate pre- and post-conditions and invariants
for advice.  Specifications in Pipa, along with aspect definitions,
are translated to JML and Java code, respectively. Pipa differs from
the proposal of Clifton and Leaves. Assertions are allowed on aspect
introductions and the \texttt{accept} expression is not provided by
Pipa. Further more, the AspectJ language is not extended to accommodate
for the definition of \texttt{observer} and \texttt{assistant} aspect
definitions.  Behavioral specifications inside aspects are translated into
JML specifications following the specification generation of execution
paths as defined in~\cite{Clifton:2002:OAP}. Pipa does not provide, nor
enforce, aspect categories.  By concentrating on AspectJ's intermediate
Java representation of aspect programs, Pipa becomes part of the AspectJ
compiler making extensions difficult and blame assignment more complex.

\PARAGRAPH{Classification system}
Rinard et al.~\cite{Rinard:2004:CSA} present a classification that is also
derived from the interaction of advice and methods.  Their focus is on
automated analysis, while our work focuses on enforcing contracts based
on annotations.  The interaction between agnostic aspects and methods can
be classified as \emph{augmentation} in their classification. The obedient
and rebellious aspects we have identified refine their \emph{combination}
class of interactions.  The \emph{narrowing} and \emph{replacement}
interactions can help in extending our work to also handle around advice
with or without proceed.  Their additional classification of scopes and
field access, can also help in extending our DbC tool to handle set-field
and get-field advice.  Automated classification to help suggest or verify
aspect categories annotation is a direction for future work.

\section{Conclusion}\label{section:conclude}

The paper discusses the intricate issues related to DbC for
AOP. The dynamic nature of aspects along with the base system's
obliviousness render existing DbC methodologies inadequate
for dealing with aspect-oriented programs. This inadequacy
unavoidably leads to erroneous error reporting and blame
assignment. An extension to DbC to address aspect-oriented
programming requires more than simple assertion validation
at each program execution point where aspect advice gets
to execute. A DbC mechanism for AOP has to address both the
execution order of contracts (on classes and aspects) as well
as the implications (if any) that must be validated between
contracts on aspects and contracts in classes. Based on these
decisions of execution order and contract implications, blame
assignment can be defined.

We provide an analysis of the possible execution order between
contracts and view the addition of advice as a behavioral
extension to the existing program.  Our analysis leads us
to a categorization of aspects into agnostic, obedient and
rebellious. Each such categorization enforces an execution
order between contracts as well as the necessary implications
that verify at runtime that the extended (through aspects)
system remains a behavioral subtype of the unextended original
system. Error reporting and blame assignment is extended to
deal with contracted aspect oriented systems. Our ideas have
been integrated in \conaj, an aspect-based DbC tool for AOP
which uses aspects to implement contracts and their runtime
validation. {\conaj} serves both as a language extension to
Java and AspectJ, but also as a case study for our own work.

Through the development and usage of {\conaj} we hope to improve
the ability to reasoning about the interactions of aspects
with other program entities including aspects themselves. We
believe that the incorporation of pre- and post-conditions on
\texttt{before} and \texttt{after} advice is a step forward
towards reasoning about aspects and their behavior.


\end{document}